\documentclass{aa}  

\usepackage{graphicx}
\usepackage{txfonts}
\usepackage{lipsum}
\usepackage{subcaption}         
\usepackage{lscape}             
\usepackage{placeins}           
                                
\usepackage{hyperref}
\hypersetup{colorlinks=true, linkcolor=blue, filecolor=blue, urlcolor=cyan, citecolor=blue}

\usepackage{natbib}

\begin{document}

   \title{A radio view on Gamma-Loud Protostars}

   \subtitle{Derivation of jet mechanical luminosity}

%
%
%

   \author{J.~{M\'endez-Gallego}\inst{1}\corrauth{jmendez@iaa.es}        
        \and C.~{Carrasco-Gonz\'alez}\inst{2}
        \and R.~{L\'opez-Coto}\inst{1}
        \and I.~{Agudo}\inst{1}
        \and S.~{Menchiari}\inst{1}
        \and R.~{Fedriani}\inst{1}
        \and E.~{de~O\~na~Wilhelmi}\inst{3}
        \and G.~Anglada\inst{1}
        }

   \institute{Instituto de Astrof\'{i}sica de Andaluc\'{i}a (IAA), CSIC, Glorieta de la Astronom\'{i}a s/n, Granada, 18008, Spain.
   \and Instituto de Radioastronom\'{i}a y Astrof\'{i}sica (IRyA), UNAM, Antigua Carretera a P\'{a}tzcuaro \#8701, 58341 Morelia (Michoac\'{a}n), Mexico.
   \and Deutsches Elektronen-Synchrotron DESY, Platanenallee 6, 15738 Zeuthen, Germany.}

   \date{Submitted {July} {27}, {2026}}

 
  \abstract
  {Gamma-Loud Protostars (GLPs) have been recently reported as Galactic hadronic accelerators whose acceleration site is situated in their protostellar jets. Theoretical and observational analysis situate radio cm luminosity as a thermal tracer of jet activity, presenting an unique opportunity to study jet properties as accelerators.}
  {We aim to develop a way to estimate the kinetic power of protostellar jets and compare their energetics with those extracted from the non-thermal $\gamma$-ray side of GLPs.}
   {We combine theoretical and phenomenological relations to estimate the jet mechanical luminosity based on the radio cm luminosity. We relate the resulting values to the non-thermal contribution of GLPs, studying its behaviour and efficiency.}
   {The derivation of the jet power successfully reproduces infrared and radio observations. The cosmic ray energy correlates to the injected mechanical energy, implying an acceleration efficiency of 1--10\%. Future radio observations are needed to find the definite accelerators and obtain reliable efficiencies.}
  {}


   \maketitle
   
   \nolinenumbers


\section{Introduction}
\label{sect:intro}

The recent discovery of Gamma-Loud Protostars \citep[GLPs;][]{2026Mendez} establishes individual young stellar objects (YSOs) as sources of Galactic cosmic rays (CRs), confirming their potential contribution to the non-thermal environment of star-forming regions. Particularly, YSOs may provide a significant population of low-energy CRs within these regions, which are among the leading candidates for Galactic PeVatrons \citep{2024deOna}. The reported $\gamma$-ray emission is consistent with hadronic origin and suggests a nearly constant particle acceleration efficiency among all identified systems, implying a direct relation between CR production and the kinetic power injected by protostellar jets, also known as mechanical luminosity. This interpretation is supported by theoretical models, predicting that shocks in protostellar jets can accelerate hadrons up to hundreds of GeV \citep{2015Padovani,2016Padovani,2021Araudo, 2026Peretti}. Additional evidence comes from individual $\gamma$-ray detections toward well-known protostellar jets such as HH~219 \citep{2022Araya}, S255 NIRS~3 \citep{2023deOna}, HH~80--81 \citep{2025Mendez}, and AFGL~490 \citep{2026Yang}.

Radio observations have historically provided a powerful tool to characterize both the thermal and non-thermal properties of protostellar jets. Thermal free-free emission produced in shocked ionized gas is closely linked to the bolometric luminosity of YSOs, correlating 3.6~cm radio continuum emission to the accretion-ejection process \citep{2018Anglada}. Similar correlations involving the energetics of CO molecular outflows \citep{1992Cabrit, 2002Beuther}, the H$_2$ infrared lines \citep{2006Caratti, 2015Caratti}, and the 3.6~cm radio continuum \citep{2018Anglada} provide important insights on the driving mechanism for outflow launching and collimation: magneto-centrifugal forces from accretion disks around YSOs eject fast and collimated material that interacts with the interstellar medium \citep{1982Blandford, 1983Pudritz}. At the same time, radio observations reveal synchrotron emission from relativistic electrons accelerated in jet shocks through linearly polarized emission \citep{2010Carrasco,2025Rodriguez}, and negative spectral indices \citep[][for individual and systematic analyses, respectively]{2016Rodriguez,2024Obonyo}. These findings support YSOs and their surroundings as particle accelerators.
 
Although the \emph{Fermi}-Large Area Telescope (LAT), a $\gamma$-ray telescope, is limited in sensitivity and angular resolution, the current GLP sample strongly suggests that particle acceleration takes place within protostellar jets \citep{2026Mendez}. However, several questions remain open.
Are the jets driven by the YSOs the dominant accelerators of these objects? Are mechanical luminosities of jets high enough to reproduce the non-thermal emission in a hadronic scenario? Do other structures such as accretion shocks or streamers also contribute? What is the efficiency needed to accelerate protons?
Radio observations are fundamental to derive proper jet mechanical luminosities and, thus, to understand the acceleration origin and measure the acceleration efficiency in a sample of hadronic accelerators.

In this work, we present a framework to address these questions: by comparing the $\gamma$-ray luminosities reported by \cite{2026Mendez} with cm~radio luminosities measured in the Very Large Array Sky Survey at 3~GHz \citep[VLASS;][]{2020Lacy} --in the Northern hemisphere--, and the SARAO MeerKAT 1.3 GHz Galactic Plane Survey \citep[hereafter SARAO;][]{2024Goedhart}, in the Southern hemisphere.
Since both distance corrected  CR output and radio continuum are expected to scale with bolometric luminosity, a correlation between these observables is anticipated. Furthermore, using the well-established relation between cm luminosity and outflow momentum rate \citep{2018Anglada}, we derive an approximate expression for the jet mechanical luminosity and use it to estimate the particle acceleration efficiency required to reproduce the observed \emph{Fermi}-LAT emission.

\section{Kinetic power traced by cm luminosity}
\label{sect:theory}

Considering the geometrical modelling for free-free emission of ionized material in  collimated protostellar outflows at cm wavelengths, Eq. (\ref{eq:reynolds}) predicts the ionized mass loss rate ($\dot{M}_{\rm ion}$) ejected in a conical jet for an optimal wavelength range, where the protostellar jet is characterized by a constant spectral index, resulting from the combination of optically thin and thick regions \citep{1986Reynolds, 2018Anglada}:
\begin{equation}
    \begin{split}
    \left( \frac{\dot{M}_{\rm ion}}{\rm 10^{-6}\, M_\odot \, yr^{-1}} \right) = 0.086 \, \left[ \left( \frac{ S_\nu \, d^2}{\rm mJy \, kpc^2} \right) \left( \frac{\nu}{\rm 10 \, GHz} \right)^{-0.6} \right]^{3/4}  \\
    \times \left( \frac{V_{\rm jet}}{\rm 200\, km\, s^{-1}} \right) \, \left( \frac{\theta_0}{\rm 30^\circ} \right)^{3/4} \left( \sin i \right)^{1/4} \, \left( \frac{T}{\rm 10^4 \, K } \right)^{-3/40},
    \end{split}
    \label{eq:reynolds}
\end{equation}
where $S_\nu \, d^2$ is the cm luminosity, $\nu$ is the observed frequency,
$V_{\rm jet}$ is the jet speed, $\theta_0$ represents the jet opening angle at the launch region, $i$ is the observed inclination, and $T$ is the temperature. For convenience, we are interested in compare Eq. (\ref{eq:reynolds}) to phenomenological relations observed at 3.6~cm. Therefore, we use $\nu = 8.3$~GHz.

\cite{2025Rota}, for example, simplified the expression above by using a classical $T$ value for protostellar jets of $10^4$~K. In the same way, \cite{2018Anglada} show that typical opening angles lie at $\theta_0 \approx (33 \pm 16)^\circ$. Omitting the weak dependence with inclination, we derive the ejected mass loss rate, considering a constant ionization fraction of $\dot{M}_{\rm ion} = 10\% \dot{M}_{\rm jet}$ \citep[e.g.,][]{2007Cabrit, 2018Anglada, 2019Fedriani}:
\begin{equation}
    \begin{split}
    \left( \frac{\dot{M}_{\rm jet}}{\rm 10^{-6}\, M_\odot \, yr^{-1}} \right) = \left( 1.0 \pm 0.4 \right) \, \left( \frac{ S_\nu \, d^2}{\rm mJy \, kpc^2} \right)^{3/4} 
    \left( \frac{V_{\rm jet}}{\rm 200\, km\, s^{-1}} \right),
    \end{split}
    \label{eq:massloss}
\end{equation}
where we adopted $\theta_0 = (33 \pm 16)^\circ$, adding uncertainties. Then, we rearrange Eq. (\ref{eq:massloss}) in order to substitute the velocity by the momentum rate of the jet ($\dot{P}_{\rm jet} \, \dot{M}_{\rm jet}^{-1} = V_{\rm jet}$, assuming $\dot{V}_{\rm jet}=0$):

\begin{equation}
    \begin{split}
    \left( \frac{\dot{M}_{\rm jet}}{\rm 10^{-6}\, M_\odot \, yr^{-1}} \right) = \left( 0.71 \pm 0.13 \right) \, \left( \frac{ S_\nu \, d^2}{\rm mJy \, kpc^2} \right)^{3/8} \\
    \times \left( \frac{\dot{P}_{\rm jet}}{\rm 10^{-4} \,  M_\odot \, yr^{-1} \, km\, s^{-1}} \right)^{1/2}.
    \end{split}
    \label{eq:momrate_reynolds}
\end{equation}

Independently, and based on observational data, \cite{2018Anglada} show a clear correlation between the momentum rate of molecular outflows ($\dot{P}_{\rm out}$) and $S_\nu \, d^2$, also calculated for 3.6~cm:

\begin{equation}
    \left( \frac{\dot{P}_{\rm out}}{\rm 10^{-4} \, M_\odot \, yr^{-1} \, km \, s^{-1}} \right) = 
     10^{1.09 \pm 0.35} \left( \frac{ S_\nu \, d^2}{\rm mJy \, kpc^2} \right)^{0.98\pm 0.08}.
    \label{eq:momrate_anglada}
\end{equation}

Equation (\ref{eq:momrate_anglada}) describes the momentum rate of the molecular outflows driven by YSOs. These massive outflows typically travel at velocities of a few to a few tens of $\rm km \, s^{-1}$, much slower than the highly collimated jets that drag them. Assuming momentum conservation, the momentum rates of the jet and the molecular outflow satisfy $\dot{P}_{\rm jet} \approx \dot{P}_{\rm out}$. Under this assumption, $\dot{P}_{\rm jet}$ in Eq. (\ref{eq:momrate_reynolds}) can be replaced by Eq. (\ref{eq:momrate_anglada}).

It is important to stress that, besides the indicated uncertainties, Eq.(\ref{eq:momrate_anglada}) has intrinsic dispersion due to the difficulties in computing $\dot{P}_{\rm out}$ because of the uncertainties in geometry, orientation, and excitation of molecular outflows.
However, we can compare the theoretical relation of Eq. (\ref{eq:momrate_reynolds}) with high expected uncertainties, and the expected phenomenology of Eq. (\ref{eq:momrate_anglada}) to get the correlation between the jet speed and the cm luminosity:

\begin{equation}
    \begin{split}
    \left( \frac{\dot{M}_{\rm jet}}{\rm 10^{-6}\, M_\odot \, yr^{-1}} \right) = 2.5_{-0.9}^{+1.4} \, \left( \frac{ S_\nu \, d^2}{\rm mJy \, kpc^2} \right)^{0.87 \pm 0.04} 
    \end{split}
    \label{eq:mjet}
\end{equation}

Equation (\ref{eq:mjet}) only depends on radio cm luminosity. The reference value is close to $\rm 10^{-6} \, M_\odot \, yr^{-1}$, which is on the order of magnitudes expected for YSOs. Besides, the mass loss rate of the jet will significantly increase with cm luminosity, as expected (see Sect. \ref{sect:validation} for details on observational validation.)

Additionally, we can employ this derivation to estimate the jet mechanical luminosity
($L_{\rm jet} = 0.5 \, \dot{P}_{\rm jet}^2 \dot{M}_{\rm jet}^{-1}$). As a result, Eq. (\ref{eq:Ljet}) combines Eq. (\ref{eq:momrate_anglada}) and (\ref{eq:mjet}):
\begin{equation}
    \left( \frac{L_{\rm jet}}{\rm L_\odot} \right) = 50_{-40}^{+120} \, \left( \frac{ S_\nu \, d^2}{\rm mJy \, kpc^2} \right)^{1.10 \pm 0.11} 
    \label{eq:Ljet}, 
\end{equation}
where the central value of 50~L$_\odot$ is highly affected by uncertainties coming from both, theoretical and phenomenological modelling. 
Therefore, the relation of the mechanical luminosity versus the radio luminosity should be validated for sufficiently large samples of observations, comparing the steepness of logarithmic measurements.

\section{Observational overview}
\label{sect:validation}

\cite{2018Anglada}, and references therein, establish two independent correlations for relating $\dot{P}_{\rm out}$ and $L_{\rm bol}$ to $S_\nu d^2$. These correlations are now well studied and have been validated over a broad range of mass and luminosity. In this work, we incorporate Eq.~(\ref{eq:reynolds}) into this correlations, since our goal is to estimate the jet power rather than the mechanical luminosity of the molecular outflows. Therefore, in this section, we compare both observational and theoretical derivations to guarantee that our results remain consistent with current observations.

\begin{figure}
    \centering
    \includegraphics[width=0.85\linewidth]{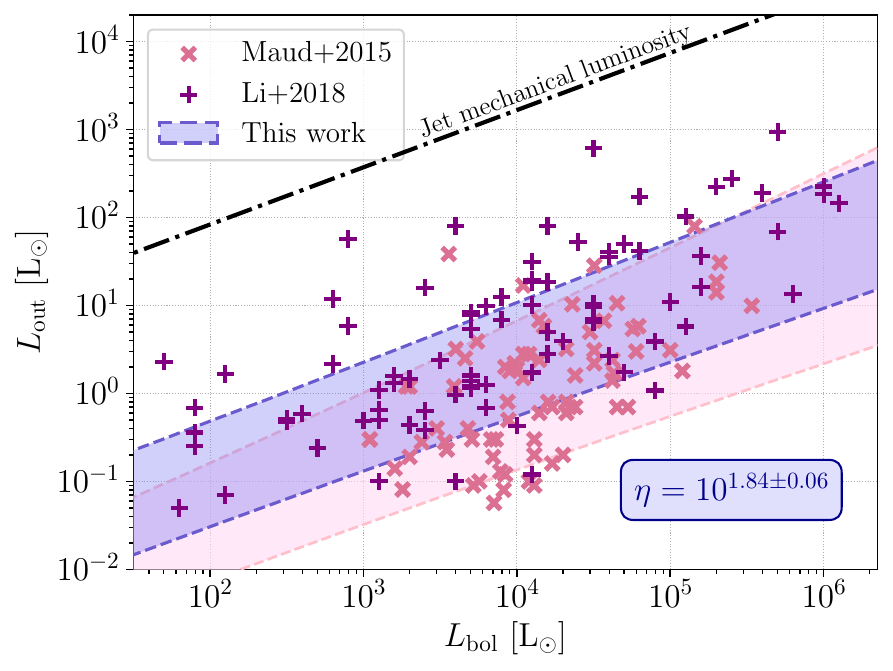}
    \caption{Mechanical luminosity of molecular outflows versus bolometric luminosity of driving YSOs. Markers show the observational data reported by \cite{2015Maud} and \cite{2018Li}. Equation (\ref{eq:Maud}) is marked with the shadowed pink area. Equation (\ref{eq:Ljet_Lbol}) (black dot-dashed line) is rescaled $\eta^{-1}$ times to match the samples of observational data (blue area).}
    \label{fig:Ljet_obs}
\end{figure}

Observational studies \citep[e.g.][]{1996Bontemps, 2014Dunham, 2015Maud, 2018Li} have revealed a clear empirical correlation between the mechanical luminosity of molecular outflows ($L_{\rm out}$) and the bolometric luminosity of the driving YSO ($L_{\rm bol}$; see Fig. \ref{fig:Ljet_obs}). For instance, \cite{2015Maud} derived the following relation:
\begin{equation}
    \left( \frac{L_{\rm out}}{\rm L_\odot} \right) = 10^{-2.92 \pm 0.62} \, \left( \frac{L_{\rm bol}}{\rm L_\odot} \right)^{0.72 \pm 0.15}. 
    \label{eq:Maud}
\end{equation}

Since Eq.~(\ref{eq:Ljet}) is written in terms of cm luminosity, we can translate it into bolometric luminosity using Eq. (28) of \cite{2018Anglada} that relates cm luminosity and $L_{\rm bol}$ for thermal protostellar jets. Note, however, that the dispersion in the $(S_\nu \, d^2)$ vs $L_{\rm bol}$ relation may add substantial errors while transforming radio fluxes into bolometric luminosities for individual cases.
\begin{equation}
    \left( \frac{ S_\nu \, d^2}{\rm mJy \, kpc^2} \right) = 10^{-1.90 \pm 0.07} \, \left( \frac{L_{\rm bol}}{\rm L_\odot} \right) ^{0.59 \pm 0.03}.
    \label{eq:Lbol_radio}
\end{equation}

By using Eq.~(\ref{eq:Lbol_radio}), we relate Eqs. (\ref{eq:mjet}) and (\ref{eq:Ljet}) to the YSO bolometric luminosity:
\begin{equation}
    \left( \frac{\dot{M}_{\rm jet}}{\rm 10^{-6} \, M_\odot \, yr^{-1}} \right) = 10^{-1.25 \pm 0.22} \, \left( \frac{L_{\rm bol}}{\rm L_\odot} \right) ^{0.51 \pm 0.03},
    \label{eq:Mjet_Lbol}
\end{equation}
\begin{equation}
    \left( \frac{L_{\rm jet}}{\rm L_\odot} \right) = 10^{-0.38 \pm 0.58} \, \left( \frac{L_{\rm bol}}{\rm L_\odot} \right) ^{0.65 \pm 0.07}.
    \label{eq:Ljet_Lbol}
\end{equation}

Equation (\ref{eq:Maud}) show that, in principle, molecular outflow will have lower mechanical luminosities than their jet counterpart. Under our assumption that $\dot{P}_{\rm jet} \approx \dot{P}_{\rm out}$, the higher velocity of the jet naturally implies a larger mechanical luminosity, since $L_{\rm jet} = 0.5\, \dot{P}_{\rm jet} V_{\rm jet}$ and, therefore, $L_{\rm jet}/L_{\rm out} \approx V_{\rm jet}/ V_{\rm out}$. 

Figure \ref{fig:Ljet_obs} shows Eq. (\ref{eq:Ljet_Lbol}) rescaled to reproduce the observational relation derived for molecular outflows. Under this scaling, the jet and outflow mechanical luminosities are related by $L_{\rm jet} = \eta L_{\rm out}$, with a best-fitting value of $\eta=10^{1.84\pm 0.06}$. The mechanical luminosity of the jet results $\sim$two orders of magnitude larger than that of the associated molecular outflow. Consequently, Eq.~(\ref{eq:Maud}) and Eq.~(\ref{eq:Ljet_Lbol}) are consistent to one another.

Assuming of $\dot{P}_{\rm jet} \approx \dot{P}_{\rm out}$, we obtain that $\eta = L_{\rm jet} / L_{\rm out} = \dot{M}_{\rm out} / \dot{M}_{\rm jet}$, where we are relating the mass loss rate ejected through the molecular outflow ($\dot{M}_{\rm out}$) with that one of the jet. We therefore transform $\dot{M}_{\rm jet}$ from Eq.~(\ref{eq:Mjet_Lbol}) into $\dot{M}_{\rm out}$ to obtain:
\begin{equation}
    \left( \frac{\dot{M}_{\rm out}}{\rm 10^{-6} \, M_\odot \, yr^{-1}} \right) = 3.9^{+2.6}_{-1.6} \, \left( \frac{L_{\rm bol}}{\rm L_\odot} \right) ^{0.51 \pm 0.03}.
    \label{eq:Mout_Lbol}
\end{equation}
For YSOs of 1 -- 1000~$\rm L_\odot$, Eq.(\ref{eq:Mout_Lbol}) returns values of the order of $\sim$10$^{-6}$~--~10$^{-4}$~$\rm M_\odot \, yr^{-1}$, consistent with observational values \citep[e.g.][]{2002Beuther}.

Since our estimates successfully reproduce the observed mass-loss rates and outflow mechanical luminosities, we conclude that the derived relations provide a reasonable framework to estimate the jet mechanical luminosity from the bolometric luminosity of the YSO.

\section{Non-thermal tracers of jet power}
\label{sect:glps}

The spectral properties of the $\gamma$-rays emitted by GLPs points towards a hadronic origin, with particles being likely accelerated by protostellar jets \citep[see][]{2026Mendez}. Typically, these hadronic accelerators are based on fast moving shocks that transform kinetical power into CR energy, with constant efficiencies of the order of $\sim$10\% \citep{2014Caprioli}, such as in supernova remnants \citep{2022Cristofari} or in young stellar clusters \citep{2024Peron}. Therefore, the ability to estimate the mechanical luminosity of protostellar jets is useful to determine the acceleration efficiency ($\varepsilon$), which will help to clarify the nature of the non-thermal particle distribution originated in GLPs. 

\subsection{Radio counterpart selection}

We initially cross-matched public catalogues from VLASS and SARAO with the observed GLP positions. To do so, we selected the nearest radio counterpart within a 5~arcmin separation (emulating the typical sizes of the field of view for radio arrays). Nevertheless, due to current all-sky radio surveys do not reach sufficient sensitivity for detecting protostellar jets, many GLPs do not have a reliable nearby radio counterpart. In these cases, nearby sources may serve at least as upper limits on the possible powering radio-source.

\begin{figure}
    \centering
    \includegraphics[width=0.85\linewidth]{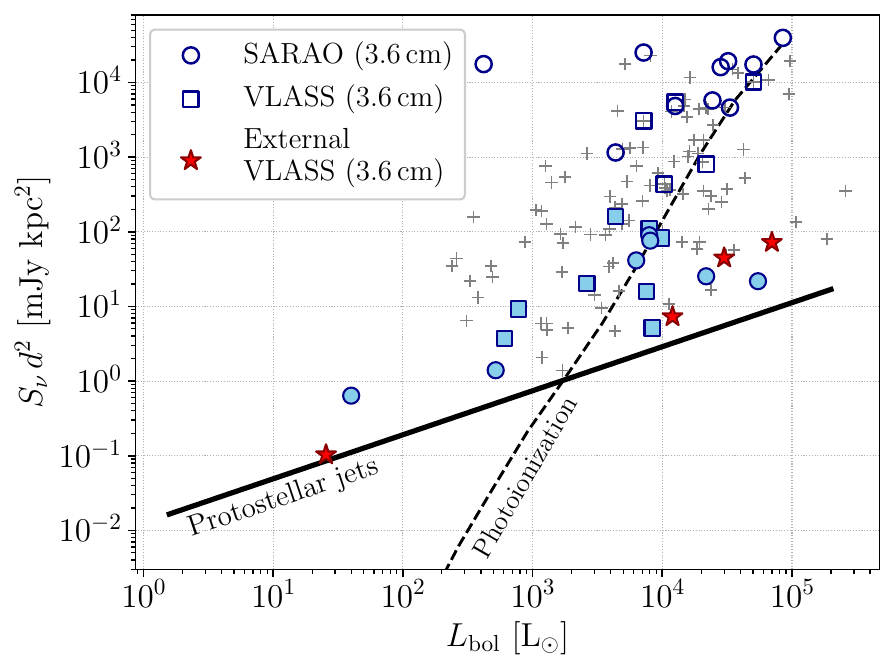}
    \caption{Radio luminosity at 3.6 cm versus bolometric luminosity for the 24 GLPs with a counterpart in VLASS or SARAO. Filled circles and squares indicate sources consistent with protostellar jets, while empty symbols follow the Ly-continuum emission for compact \ion{H}{ii} regions. Red stars mark the external GLP associations reported by \cite{2026Mendez}, all of them behaving as protostellar jets. Grey crosses show the radio counterparts identified within 1~arcmin of the YSO positions. Solid black line marks the relation expected for protostellar jets and dashed line is the relation for photoionized sources, both extracted from \cite{2018Anglada}.}
    \label{fig:YSO}
\end{figure}

In order to test the reliability of the selected radio counterparts, Fig. \ref{fig:YSO} shows the cm luminosities versus $L_{\rm bol}$. To keep consistency with \cite{2018Anglada}, we scaled the VLASS S-band (3~GHz) and SARAO 1.3~GHz corresponding fluxes into a wavelength of 3.6~cm, using a spectral index of 0.6 for the entire thermal radio jet. 

It is well established that protostellar jets follow a power-law relation between bolometric luminosity, used as a proxy for the accretion/ejection power, and radio continuum emission. The latter arises from free-free radiation produced in shock-ionized material within the outflow. However, once the central protostar becomes sufficiently massive and evolved to emit ionizing UV photons, the free-free emission becomes dominated by photoionization rather than shock ionization. This transition typically indicates a decline in accretion activity, at which point the protostellar jet activity decays.

In a similar way, we cross-matched all YSO positions compiled in the Red MSX Source survey \citep{2013Lumsden} with SARAO and VLASS, finding that only $\sim$30\% of YSOs within the VLASS coverage area have a radio counterpart within 1~arcmin. For SARAO, no close counterparts were identified.
Moreover, the grey crosses in Fig.~\ref{fig:YSO} show that the detected YSOs predominantly follow the behaviour expected for photoionized regions (\ion{H}{ii} candidates), indicating that these radio surveys do not have sufficient sensitivity to properly sample our GLP population. Nevertheless, \ion{H}{ii} candidates and protostellar jets relation versus cm luminosities can still be identified among our cross-matched sample of GLPs. Our sample of data points usually lies slightly over the expected relations, indicating a maximum radio-brightness for our GLPs and highlighting their behaviour in the plot.

Figure \ref{fig:YSO} shows that many GLPs lie directly along the relation reported by \cite{2018Anglada} for protostellar jets (filled blue markers), and are clearly separated from compact \ion{H}{ii} regions associated with photoionization (empty markers). In these last cases, if nearby protostellar jets are present around a compact \ion{H}{ii} region, the radio surveys may miss the much fainter jet emission, thereby hiding the true origin of the GLP. Only dedicated, high-sensitivity radio observations would allow a reliable identification of the powering sources.
In any case, based on this figure, it is evident that GLPs are associated with YSOs in which forming stars can either launch powerful jets or ionize their surrounding interstellar medium. Sources whose cm luminosities is $>$50 times above the relation for protostellar jets are classified as candidates for \ion{H}{ii} regions.

\subsection{Estimation of acceleration efficiencies}

\begin{figure}
    \centering
    \includegraphics[width=0.85\linewidth]{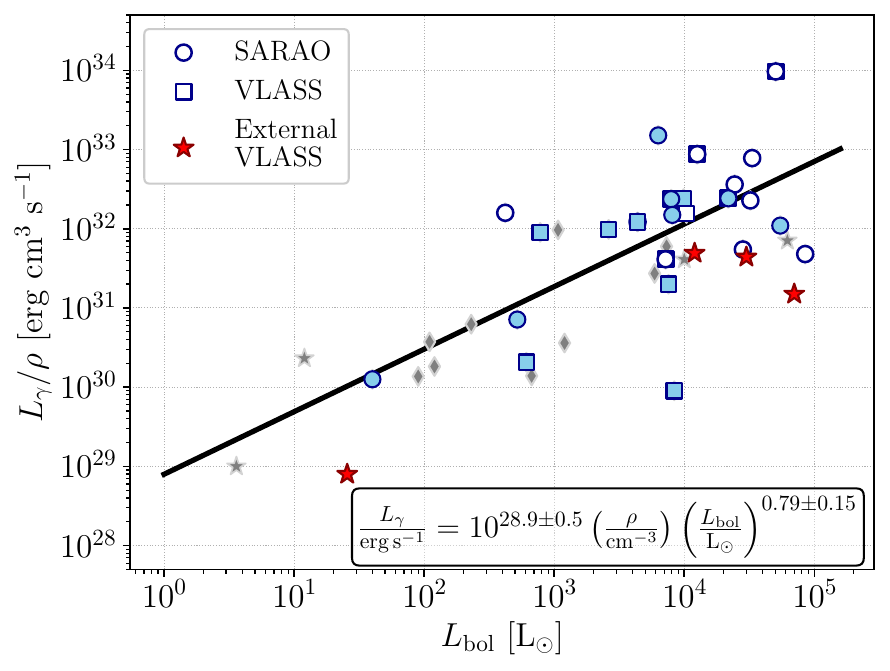}
    \caption{CR output ($L_\gamma / \rho$) versus YSO bolometric luminosity. Marker styles follow the same notation as in Fig. \ref{fig:YSO}. Grey markers show the position of GLPs (diamonds) and external sources (stars) that were not identified in VLASS or SARAO. Solid line show the correlation found by \cite{2026Mendez}.}
    \label{fig:mendez26}
\end{figure}

\cite{2026Mendez} presented a correlation between the YSO luminosity and CR output, quantified as the $\gamma$-ray luminosity ($L_\gamma$; $>$100~MeV) normalized by the ambient density ($\rho$). Figure \ref{fig:mendez26} shows no significant differences between the populations previously identified as protostellar jet and \ion{H}{ii} candidates in Fig. \ref{fig:YSO}. This supports the possibility that other YSOs powering jets might be outshined by the strong emission emanating from compact \ion{H}{ii} regions. It is also important to highlight that the relation shown in Fig. \ref{fig:mendez26} and Eq. (\ref{eq:Ljet_Lbol}) have consistent power-law indexes. This points towards CR production scaling proportionally to the mechanical luminosity of jets.

\begin{figure}
    \centering
    \includegraphics[width=0.95\linewidth]{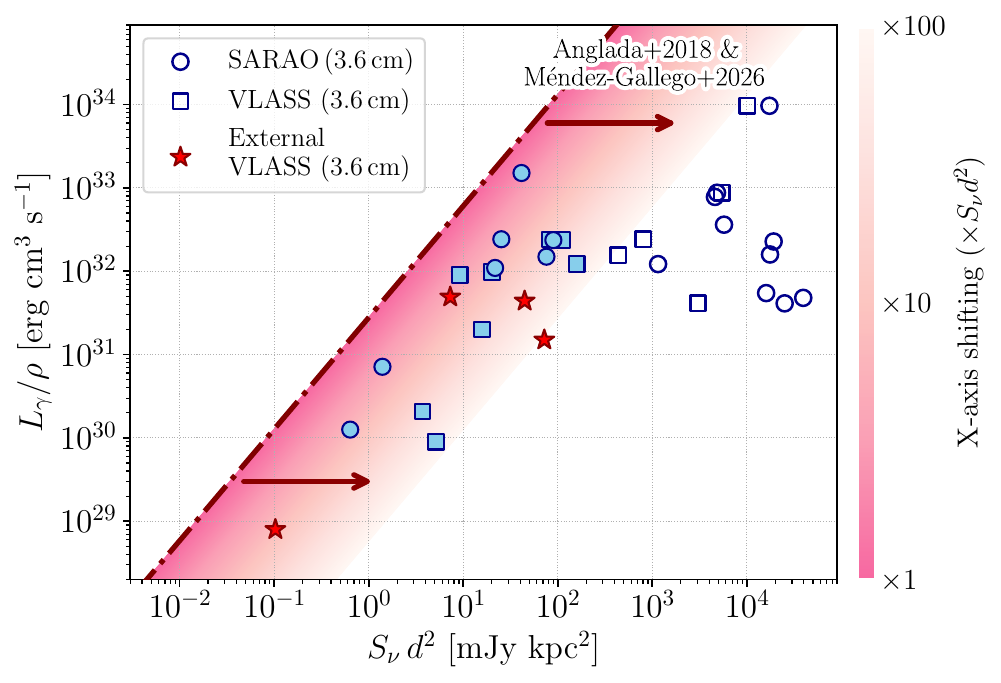}
    \caption{CR output versus cm luminosity at 3.6~cm. Marker styles follow the same notation as in Fig. \ref{fig:YSO}. The combination of relations from Figs. \ref{fig:YSO} and \ref{fig:mendez26} is a dot-dashed line. The colorbar indicates the x-axis shift of the dot-dashed line.}
    \label{fig:gamma_radio}
\end{figure}

Based on the solid synergies that cm luminosity and CR production have, Fig.~\ref{fig:gamma_radio} combines \cite{2018Anglada} and \cite{2026Mendez} relations to unveil the behaviour of CR output compared to the radio luminosity. 
As expected, the \ion{H}{ii} candidates are uncorrelated with the expected relation between radio jets and GLP's high-energy luminosity. This can be interpreted as a result of the $S_\nu \, d^2$ excess shown in Fig.~\ref{fig:YSO} versus their $L_{\rm bol}$. 
For filled markers, in contrast, the observed slope is consistent with the expectations. However, a clear shifting to the right side of the plot is observed, possibly indicating a radio flux excess or a lack of $L_\gamma$. Based on the general bias found in Fig.~\ref{fig:YSO}, dedicated radio observations at the correct wavelengths and with accurate sensitivity would be needed to understand the nature of the shifting.

\begin{equation}
    E_{\rm CR} \approx \varepsilon \, \tau \, L_{\rm jet}.
    \label{eq:Ecr}
\end{equation}

Equation (\ref{eq:Ecr}) provides a simple model for the conversion of the jet mechanical luminosity into CRs through diffusive shock acceleration. Here, $E_{\rm CR}$ represents the energy injected into a non-thermal population of relativistic protons. Although the lifetime of protostellar jets ($\tau$) may vary from source to source, we adopt a representative value of $\tau = 10^5$~yr \citep[see, e.g.,][]{2014Frank} for computing the total kinetic energy ejected by protostellar jets. Using archival data from VLASS and SARAO, together with the relation derived in Eq. (\ref{eq:Ljet}), we estimate the $\varepsilon$ required to account for the observed $\gamma$-ray emission. Our approach implicitly assumes that GLPs do not host unusually energetic jets, i.e. that their jets are broadly consistent with the energetics of the thermal jets observed in other `Gamma-Quiet' YSOs in which $\gamma$-ray emission has not been detected. Note however that sensitive radio observations are required, since compiled data points from Figs. \ref{fig:YSO} and \ref{fig:gamma_radio} are subject to observational biases.

\begin{figure}
    \centering
    \includegraphics[width=0.85\linewidth]{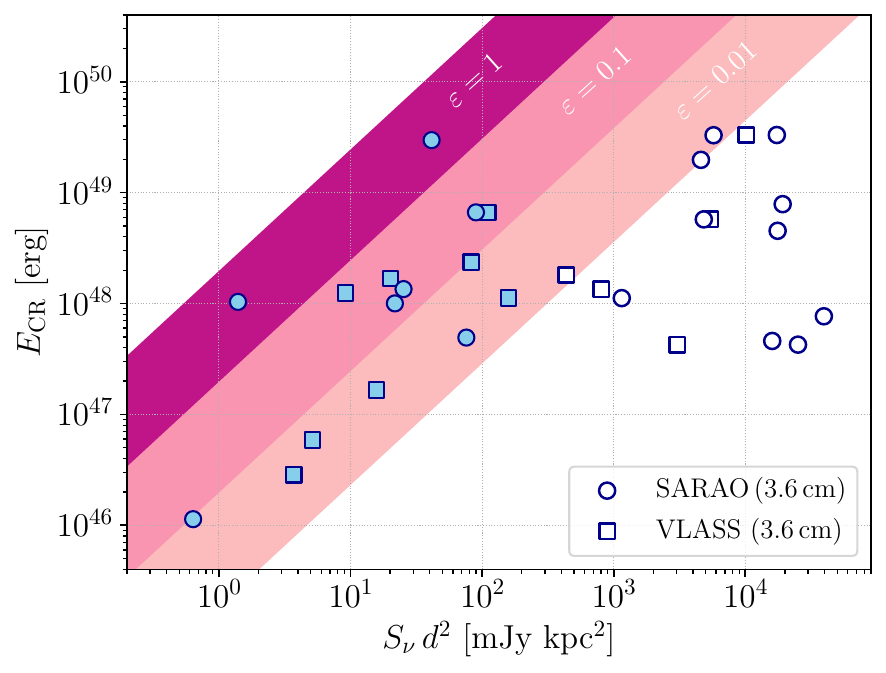}
    \caption{Non-thermal CR energy calculated for a hadronic scenario versus radio cm luminosity. Markers follow the same notation as in Fig.~\ref{fig:YSO}. Colored areas show the expected CR energy computed using Eq.~(\ref{eq:Ljet}) for different efficiencies.}
    \label{fig:efficiency}
\end{figure}

Figure~\ref{fig:efficiency} shows the hadronic non-thermal energy estimated to reproduce the observed $\gamma$-ray emission under a purely hadronic scenario \citep{2026Mendez}. If the jets are indeed the particle accelerators, a fraction of their kinetic energy must be transferred to CRs. As discussed by \cite{2026Mendez}, the estimates of $E_{\rm CR}$ are affected by substantial dispersion. Therefore, the individual points should not be interpreted in isolation, but rather as part of the overall population.

Note that, since the jet mechanical luminosity -- and thus, the ejected kinetic energy-- depends on the cm luminosity, the x-axis of Fig. \ref{fig:efficiency} is a proxy to the protostellar jet energetics. In the figure, no protostellar jet is above the maximum possible efficiency ($\varepsilon = 1$). Based on VLASS and SARAO data, and ignoring the observational biases, the acceleration efficiency seems to be in the range of 1--10\%, consistent with \cite{2014Caprioli}. Consequently, radio data suggest that GLPs are powered by protostellar jets able to accelerate hadrons via diffusive shock acceleration.



\section{Conclusion}
\label{sect:conclusion}
We have presented a method to estimate the mechanical power of jets launched by YSOs. Our results support the scenario in which protostellar jets are the primary particle accelerators within GLP environments, without requiring additional acceleration sites or sources. However, as current all-sky radio surveys are biased toward the brightest radio emitters, they preferentially detect photoionized regions rather than the much fainter protostellar jets, which are expected to be responsible for the observed $\gamma$-ray emission.
The main conclusions of this work are presented as follows:
\begin{itemize}
    \item Observational and theoretical models provide a well-studied framework in which the jet mechanical power can be estimated from radio-continuum emission, see Eqs. (\ref{eq:Ljet}) and (\ref{eq:Ljet_Lbol}). In this work, we used infrared and radio -wavelength surveys to show that both relations are self-consistent and can reproduce the observations.
    \item GLPs high-energy emission traces the kinetic output of protostellar jets, linking the $\gamma$-ray emission to the jet mechanical luminosity. In this work, we show that GLPs may transform the energy ejected from protostellar jets into hadronic CRs via diffusive shock acceleration.
    \item A significant portion of the current GLP sample is behaving as possible compact \ion{H}{ii} regions, hinting the presence of hidden protostellar jets. Dedicated radio observations are necessary to unveil the definite accelerators of that regions.
    \item The formulae here derived allow the first systematic derivation of acceleration efficiency in a large sample of Galactic hadronic accelerators. Although the lack of observations in the radio band only allow us to obtain rough estimates, we hint an average value of $\sim$1--10\%, in good agreement with theory.
\end{itemize}

\begin{acknowledgements}
    J.M.-G. acknowledges financial support from the FPI-Severo Ochoa grant CEX2021-001131-S-20-6, PRE2022-103386 funded by MICIU/AEI/ 10.13039/501100011033 and ESF+.
    J.M.-G., R.L.-C., S.M. and I.A. also acknowledge funding support from the Spanish ``Ministerio de Ciencia e Innovaci\'on'' through grant PID2022-139117NB-C44. 
    J.M.-G., R.L.-C., S.M., R.F., I.A., and G.A. acknowledge financial support from the Severo Ochoa grant CEX2021-001131-S funded by MCIN/AEI/ 10.13039/501100011033.
    C.C.-G. acknowledges support from DGAPA-PAPIIT project IG101224.
    R. L.-C. and S.M. also acknowledge the financial support through grant CNS2023-144504 funded by MICIU/AEI/ 10.13039/501100011033 and by the ``European Union NextGenerationEU/PRTR''. 
    EdOW acknowledges the support of DESY (Zeuthen), a member of the Helmholtz Association HGF.
    R.F. and G.A. acknowledges financial support from PID2023-146295NB-I00 funded by MCIN/AEI/ 10.13039/501100011033 and by ``European Union NextGenerationEU/PRTR'' and by ERDF/EU.
      
\end{acknowledgements}


\bibliographystyle{aa} 
\bibliography{references}

@ARTICLE{2026Mendez,
       author = {{M{\'e}ndez-Gallego}, Javier and {L{\'o}pez-Coto}, Rub{\'e}n and {de O{\~n}a Wilhelmi}, Emma and {Menchiari}, Stefano and {Agudo}, Iv{\'a}n and {Fedriani}, Rub{\'e}n},
        title = "{Evidence for protostellar jets as a population of hadronic gamma-ray sources}",
      journal = {Nature Astronomy},
         year = 2026,
        month = jun,
          doi = {10.1038/s41550-026-02886-7},
archivePrefix = {arXiv},
       eprint = {2606.19445},
 primaryClass = {astro-ph.HE},
       adsurl = {https://ui.adsabs.harvard.edu/abs/2026NatAs.tmp..133M}
}

@ARTICLE{2024deOna,
       author = {{de O{\~n}a Wilhelmi}, Emma and {L{\'o}pez-Coto}, Ruben and {Aharonian}, Felix and {Amato}, Elena and {Cao}, Zhen and {Gabici}, Stefano and {Hinton}, Jim},
        title = "{The hunt for PeVatrons as the origin of the most energetic photons observed in the Galaxy}",
      journal = {Nature Astronomy},
         year = 2024,
        month = apr,
       volume = {8},
        pages = {425-431},
          doi = {10.1038/s41550-024-02224-9},
archivePrefix = {arXiv},
       eprint = {2404.16591},
 primaryClass = {astro-ph.HE},
       adsurl = {https://ui.adsabs.harvard.edu/abs/2024NatAs...8..425D}
}

@ARTICLE{2015Padovani,
       author = {{Padovani}, M. and {Hennebelle}, P. and {Marcowith}, A. and {Ferri{\`e}re}, K.},
        title = "{Cosmic-ray acceleration in young protostars}",
      journal = {\aap},
         year = 2015,
        month = oct,
       volume = {582},
          eid = {L13},
        pages = {L13},
          doi = {10.1051/0004-6361/201526874},
archivePrefix = {arXiv},
       eprint = {1509.06416},
 primaryClass = {astro-ph.SR},
       adsurl = {https://ui.adsabs.harvard.edu/abs/2015A&A...582L..13P}
}

@ARTICLE{2016Padovani,
       author = {{Padovani}, M. and {Marcowith}, A. and {Hennebelle}, P. and {Ferri{\`e}re}, K.},
        title = "{Protostars: Forges of cosmic rays?}",
      journal = {\aap},
         year = 2016,
        month = may,
       volume = {590},
          eid = {A8},
        pages = {A8},
          doi = {10.1051/0004-6361/201628221},
archivePrefix = {arXiv},
       eprint = {1602.08495},
 primaryClass = {astro-ph.HE},
       adsurl = {https://ui.adsabs.harvard.edu/abs/2016A&A...590A...8P}
}

@ARTICLE{2021Araudo,
       author = {{Araudo}, Anabella T. and {Padovani}, Marco and {Marcowith}, Alexandre},
        title = "{Particle acceleration and magnetic field amplification in massive young stellar object jets}",
      journal = {\mnras},
         year = 2021,
        month = jun,
       volume = {504},
       number = {2},
        pages = {2405-2419},
          doi = {10.1093/mnras/stab635},
archivePrefix = {arXiv},
       eprint = {2102.11583},
 primaryClass = {astro-ph.HE},
       adsurl = {https://ui.adsabs.harvard.edu/abs/2021MNRAS.504.2405A}
}

@ARTICLE{2026Yang,
       author = {{Yang}, Li-Nuo and {Tang}, Sheng and {Tam}, Pak-Hin Thomas},
        title = "{GeV {\ensuremath{\gamma}}-ray emission in the low-mass star-forming region AFGL 490}",
      journal = {\mnras},
         year = 2026,
        month = jun,
       volume = {549},
       number = {1},
          eid = {stag896},
        pages = {stag896},
          doi = {10.1093/mnras/stag896},
archivePrefix = {arXiv},
       eprint = {2605.15489},
 primaryClass = {astro-ph.HE},
       adsurl = {https://ui.adsabs.harvard.edu/abs/2026MNRAS.549ag896Y}
}

@ARTICLE{2022Araya,
       author = {{Araya}, M. and {Guti{\'e}rrez}, L. and {Kerby}, S.},
        title = "{Probing the origin of 4FGL J0822.8-4207: cosmic ray illumination from the SNR Puppis A and the Herbig-Haro object HH219}",
      journal = {\mnras},
         year = 2022,
        month = feb,
       volume = {510},
       number = {2},
        pages = {2277-2285},
          doi = {10.1093/mnras/stab3340},
archivePrefix = {arXiv},
       eprint = {2111.09427},
 primaryClass = {astro-ph.HE},
       adsurl = {https://ui.adsabs.harvard.edu/abs/2022MNRAS.510.2277A}
}

@ARTICLE{2023deOna,
       author = {{de O{\~n}a Wilhelmi}, Emma and {L{\'o}pez-Coto}, Rub{\'e}n and {Su}, Yang},
        title = "{High-energy gamma-ray emission powered by a young protostar: the case of S255 NIRS 3}",
      journal = {\mnras},
         year = 2023,
        month = jul,
       volume = {523},
       number = {1},
        pages = {105-110},
          doi = {10.1093/mnras/stad1413},
archivePrefix = {arXiv},
       eprint = {2305.04571},
 primaryClass = {astro-ph.HE},
       adsurl = {https://ui.adsabs.harvard.edu/abs/2023MNRAS.523..105D}
}

@ARTICLE{2025Mendez,
       author = {{M{\'e}ndez-Gallego}, J. and {L{\'o}pez-Coto}, R. and {de O{\~n}a Wilhelmi}, E. and {Fedriani}, R. and {Otero-Santos}, J. and {Cant{\"u}rk}, Y.},
        title = "{Exploring the capability of the HH 80-81 protostellar jet to accelerate relativistic particles}",
      journal = {\aap},
         year = 2025,
        month = mar,
       volume = {695},
          eid = {A11},
        pages = {A11},
          doi = {10.1051/0004-6361/202452473},
archivePrefix = {arXiv},
       eprint = {2502.01261},
 primaryClass = {astro-ph.HE},
       adsurl = {https://ui.adsabs.harvard.edu/abs/2025A&A...695A..11M}
}

@ARTICLE{2018Anglada,
       author = {{Anglada}, Guillem and {Rodr{\'\i}guez}, Luis F. and {Carrasco-Gonz{\'a}lez}, Carlos},
        title = "{Radio jets from young stellar objects}",
      journal = {\aapr},
         year = 2018,
        month = jun,
       volume = {26},
       number = {1},
          eid = {3},
        pages = {3},
          doi = {10.1007/s00159-018-0107-z},
archivePrefix = {arXiv},
       eprint = {1806.06444},
 primaryClass = {astro-ph.SR},
       adsurl = {https://ui.adsabs.harvard.edu/abs/2018A&ARv..26....3A}
}

@ARTICLE{2006Caratti,
       author = {{Caratti o Garatti}, A. and {Giannini}, T. and {Nisini}, B. and {Lorenzetti}, D.},
        title = "{H$_{2}$ active jets in the near IR as a probe of protostellar evolution}",
      journal = {\aap},
         year = 2006,
        month = apr,
       volume = {449},
       number = {3},
        pages = {1077-1088},
          doi = {10.1051/0004-6361:20054313},
       adsurl = {https://ui.adsabs.harvard.edu/abs/2006A&A...449.1077C}
}

@ARTICLE{2015Caratti,
       author = {{Caratti o Garatti}, A. and {Stecklum}, B. and {Linz}, H. and {Garcia Lopez}, R. and {Sanna}, A.},
        title = "{A near-infrared spectroscopic survey of massive jets towards extended green objects}",
      journal = {\aap},
         year = 2015,
        month = jan,
       volume = {573},
          eid = {A82},
        pages = {A82},
          doi = {10.1051/0004-6361/201423992},
archivePrefix = {arXiv},
       eprint = {1410.4041},
 primaryClass = {astro-ph.SR},
       adsurl = {https://ui.adsabs.harvard.edu/abs/2015A&A...573A..82C}
}

@ARTICLE{1982Blandford,
       author = {{Blandford}, R.~D. and {Payne}, D.~G.},
        title = "{Hydromagnetic flows from accretion disks and the production of radio jets.}",
      journal = {\mnras},
         year = 1982,
        month = jun,
       volume = {199},
        pages = {883-903},
          doi = {10.1093/mnras/199.4.883},
       adsurl = {https://ui.adsabs.harvard.edu/abs/1982MNRAS.199..883B}
}

@ARTICLE{1983Pudritz,
       author = {{Pudritz}, R.~E. and {Norman}, C.~A.},
        title = "{Centrifugally driven winds from contracting molecular disks}",
      journal = {\apj},
         year = 1983,
        month = nov,
       volume = {274},
        pages = {677-697},
          doi = {10.1086/161481},
       adsurl = {https://ui.adsabs.harvard.edu/abs/1983ApJ...274..677P}
}

@ARTICLE{2010Carrasco,
       author = {{Carrasco-Gonz{\'a}lez}, Carlos and {Rodr{\'\i}guez}, Luis F. and {Anglada}, Guillem and {Mart{\'\i}}, Josep and {Torrelles}, Jos{\'e} M. and {Osorio}, Mayra},
        title = "{A Magnetized Jet from a Massive Protostar}",
      journal = {Science},
         year = 2010,
        month = nov,
       volume = {330},
       number = {6008},
        pages = {1209},
          doi = {10.1126/science.1195589},
archivePrefix = {arXiv},
       eprint = {1011.6254},
 primaryClass = {astro-ph.GA},
       adsurl = {https://ui.adsabs.harvard.edu/abs/2010Sci...330.1209C}
}

@ARTICLE{2016Rodriguez,
       author = {{Rodr{\'\i}guez-Kamenetzky}, Adriana and {Carrasco-Gonz{\'a}lez}, Carlos and {Araudo}, Anabella and {Torrelles}, Jos{\'e} M. and {Anglada}, Guillem and {Mart{\'\i}}, Josep and {Rodr{\'\i}guez}, Luis F. and {Valotto}, Carlos},
        title = "{Investigating Particle Acceleration in Protostellar Jets: The Triple Radio Continuum Source in Serpens}",
      journal = {\apj},
         year = 2016,
        month = feb,
       volume = {818},
       number = {1},
          eid = {27},
        pages = {27},
          doi = {10.3847/0004-637X/818/1/27},
archivePrefix = {arXiv},
       eprint = {1512.02980},
 primaryClass = {astro-ph.SR},
       adsurl = {https://ui.adsabs.harvard.edu/abs/2016ApJ...818...27R}
}

@ARTICLE{2024Obonyo,
       author = {{Obonyo}, W.~O. and {Hoare}, M.~G. and {Lumsden}, S.~L. and {Thompson}, M.~A. and {Chibueze}, J.~O. and {Cotton}, W.~D. and {Rigby}, A. and {Leto}, P. and {Trigilio}, C. and {Williams}, G.~M.},
        title = "{Non-thermal radio emission from massive protostars in the SARAO MeerKAT Galactic Plane Survey}",
      journal = {\mnras},
         year = 2024,
        month = oct,
       volume = {533},
       number = {4},
        pages = {3862-3877},
          doi = {10.1093/mnras/stae2020},
archivePrefix = {arXiv},
       eprint = {2505.01683},
 primaryClass = {astro-ph.GA},
       adsurl = {https://ui.adsabs.harvard.edu/abs/2024MNRAS.533.3862O}
}

@ARTICLE{2025Rodriguez,
       author = {{Rodr{\'\i}guez-Kamenetzky}, A. and {Pasetto}, A. and {Carrasco-Gonz{\'a}lez}, C. and {Rodr{\'\i}guez}, L.~F. and {G{\'o}mez}, J.~L. and {Anglada}, G. and {Torrelles}, J.~M. and {Gomes}, N.~R.~C. and {Vig}, S. and {Mart{\'\i}}, J.},
        title = "{Helical Magnetic Field in a Massive Protostellar Jet}",
      journal = {\apjl},
         year = 2025,
        month = jan,
       volume = {978},
       number = {2},
          eid = {L31},
        pages = {L31},
          doi = {10.3847/2041-8213/ad9b26},
archivePrefix = {arXiv},
       eprint = {2501.07622},
 primaryClass = {astro-ph.GA},
       adsurl = {https://ui.adsabs.harvard.edu/abs/2025ApJ...978L..31R}
}

@ARTICLE{2024Goedhart,
       author = {{Goedhart}, S. and {Cotton}, W.~D. and {Camilo}, F. and {Thompson}, M.~A. and {Umana}, G. and {Bietenholz}, M. and {Woudt}, P.~A. and {Anderson}, L.~D. and {Bordiu}, C. and {Buckley}, D.~A.~H. and et al.},
        title = "{The SARAO MeerKAT 1.3 GHz Galactic Plane Survey}",
      journal = {\mnras},
         year = 2024,
        month = jun,
       volume = {531},
       number = {1},
        pages = {649-681},
          doi = {10.1093/mnras/stae1166},
archivePrefix = {arXiv},
       eprint = {2312.07275},
 primaryClass = {astro-ph.GA},
       adsurl = {https://ui.adsabs.harvard.edu/abs/2024MNRAS.531..649G}
}

@ARTICLE{2020Lacy,
       author = {{Lacy}, M. and {Baum}, S.~A. and {Chandler}, C.~J. and {Chatterjee}, S. and {Clarke}, T.~E. and {Deustua}, S. and {English}, J. and {Farnes}, J. and {Gaensler}, B.~M. and {Gugliucci}, N. and et al.},
        title = "{The Karl G. Jansky Very Large Array Sky Survey (VLASS). Science Case and Survey Design}",
      journal = {\pasp},
         year = 2020,
        month = mar,
       volume = {132},
       number = {1009},
          eid = {035001},
        pages = {035001},
          doi = {10.1088/1538-3873/ab63eb},
archivePrefix = {arXiv},
       eprint = {1907.01981},
 primaryClass = {astro-ph.IM},
       adsurl = {https://ui.adsabs.harvard.edu/abs/2020PASP..132c5001L}
}

@ARTICLE{1986Reynolds,
       author = {{Reynolds}, S.~P.},
        title = "{Continuum Spectra of Collimated, Ionized Stellar Winds}",
      journal = {\apj},
         year = 1986,
        month = may,
       volume = {304},
        pages = {713},
          doi = {10.1086/164209},
       adsurl = {https://ui.adsabs.harvard.edu/abs/1986ApJ...304..713R}
}

@ARTICLE{2025Rota,
       author = {{Rota}, A.~A. and {van der Marel}, N. and {Garufi}, A. and {Carrasco-Gonz{\'a}lez}, C. and {Macias}, E. and {Pascucci}, I. and {Sellek}, A. and {Testi}, L. and {Isella}, A. and {Facchini}, S.},
        title = "{A correlation between accretion and outflow rates for class II young stellar objects with full and transition disks}",
      journal = {\aap},
         year = 2025,
        month = aug,
       volume = {700},
          eid = {A32},
        pages = {A32},
          doi = {10.1051/0004-6361/202554259},
archivePrefix = {arXiv},
       eprint = {2505.16586},
 primaryClass = {astro-ph.EP},
       adsurl = {https://ui.adsabs.harvard.edu/abs/2025A&A...700A..32R}
}

@ARTICLE{2019Fedriani,
       author = {{Fedriani}, R. and {Caratti o Garatti}, A. and {Purser}, S.~J.~D. and {Sanna}, A. and {Tan}, J.~C. and {Garcia-Lopez}, R. and {Ray}, T.~P. and {Coffey}, D. and {Stecklum}, B. and {Hoare}, M.},
        title = "{Measuring the ionisation fraction in a jet from a massive protostar}",
      journal = {Nature Communications},
         year = 2019,
        month = aug,
       volume = {10},
          eid = {3630},
        pages = {3630},
          doi = {10.1038/s41467-019-11595-x},
archivePrefix = {arXiv},
       eprint = {1908.05346},
 primaryClass = {astro-ph.SR},
       adsurl = {https://ui.adsabs.harvard.edu/abs/2019NatCo..10.3630F}
}

@ARTICLE{2015Maud,
       author = {{Maud}, L.~T. and {Moore}, T.~J.~T. and {Lumsden}, S.~L. and {Mottram}, J.~C. and {Urquhart}, J.~S. and {Hoare}, M.~G.},
        title = "{A distance-limited sample of massive molecular outflows}",
      journal = {\mnras},
         year = 2015,
        month = oct,
       volume = {453},
       number = {1},
        pages = {645-665},
          doi = {10.1093/mnras/stv1635},
archivePrefix = {arXiv},
       eprint = {1509.00199},
 primaryClass = {astro-ph.SR},
       adsurl = {https://ui.adsabs.harvard.edu/abs/2015MNRAS.453..645M}
}

@ARTICLE{2018Li,
       author = {{Li}, Qiang and {Zhou}, Jianjun and {Esimbek}, Jarken and {He}, Yuxin and {Baan}, W.~A. and {Li}, Dalei and {Wu}, Gang and {Tang}, Xindi and {Ji}, Weiguang and {Zhexeray}, Dauren},
        title = "{High-mass Outflows Identified from COHRS CO (3-2) Survey}",
      journal = {\apj},
         year = 2018,
        month = nov,
       volume = {867},
       number = {2},
          eid = {167},
        pages = {167},
          doi = {10.3847/1538-4357/aae2b8},
archivePrefix = {arXiv},
       eprint = {1809.08739},
 primaryClass = {astro-ph.GA},
       adsurl = {https://ui.adsabs.harvard.edu/abs/2018ApJ...867..167L}
}

@ARTICLE{2024Peron,
       author = {{Peron}, Giada and {Casanova}, Sabrina and {Gabici}, Stefano and {Baghmanyan}, Vardan and {Aharonian}, Felix},
        title = "{The contribution of winds from star clusters to the Galactic cosmic-ray population}",
      journal = {Nature Astronomy},
         year = 2024,
        month = apr,
       volume = {8},
        pages = {530-537},
          doi = {10.1038/s41550-023-02168-6},
archivePrefix = {arXiv},
       eprint = {2407.07509},
 primaryClass = {astro-ph.HE},
       adsurl = {https://ui.adsabs.harvard.edu/abs/2024NatAs...8..530P}
}

@ARTICLE{2013Lumsden,
       author = {{Lumsden}, S.~L. and {Hoare}, M.~G. and {Urquhart}, J.~S. and {Oudmaijer}, R.~D. and {Davies}, B. and {Mottram}, J.~C. and {Cooper}, H.~D.~B. and {Moore}, T.~J.~T.},
        title = "{The Red MSX Source Survey: The Massive Young Stellar Population of Our Galaxy}",
      journal = {\apjs},
         year = 2013,
        month = sep,
       volume = {208},
       number = {1},
          eid = {11},
        pages = {11},
          doi = {10.1088/0067-0049/208/1/11},
archivePrefix = {arXiv},
       eprint = {1308.0134},
 primaryClass = {astro-ph.GA},
       adsurl = {https://ui.adsabs.harvard.edu/abs/2013ApJS..208...11L}
}

@INPROCEEDINGS{2014Frank,
       author = {{Frank}, A. and {Ray}, T.~P. and {Cabrit}, S. and {Hartigan}, P. and {Arce}, H.~G. and {Bacciotti}, F. and {Bally}, J. and {Benisty}, M. and {Eisl{\"o}ffel}, J. and {G{\"u}del}, M. and et al.},
        title = "{Jets and Outflows from Star to Cloud: Observations Confront Theory}",
    booktitle = {Protostars and Planets VI},
         year = 2014,
       editor = {{Beuther}, Henrik and {Klessen}, Ralf S. and {Dullemond}, Cornelis P. and {Henning}, Thomas},
        month = jan,
        pages = {451-474},
          doi = {10.2458/azu_uapress_9780816531240-ch020},
archivePrefix = {arXiv},
       eprint = {1402.3553},
 primaryClass = {astro-ph.SR},
       adsurl = {https://ui.adsabs.harvard.edu/abs/2014prpl.conf..451F}
}

@ARTICLE{2022Cristofari,
       author = {{Cristofari}, Pierre and {Blasi}, Pasquale and {Caprioli}, Damiano},
        title = "{Microphysics of Diffusive Shock Acceleration: Impact on the Spectrum of Accelerated Particles}",
      journal = {\apj},
         year = 2022,
        month = may,
       volume = {930},
       number = {1},
          eid = {28},
        pages = {28},
          doi = {10.3847/1538-4357/ac6182},
archivePrefix = {arXiv},
       eprint = {2203.15624},
 primaryClass = {astro-ph.HE},
       adsurl = {https://ui.adsabs.harvard.edu/abs/2022ApJ...930...28C}
}

@ARTICLE{2026Peretti,
       author = {{Peretti}, Enrico and {Amato}, Elena and {Cerri}, Silvio Sergio and {Morlino}, Giovanni and {Perfetta Pullano}, Letizia and {Recchia}, Sarah},
        title = "{Particle acceleration at recollimation shocks in sub-relativistic jets A model for jets in Seyfert Galaxies, Microquasars and Protostellar Systems}",
      journal = {arXiv e-prints},
         year = 2026,
        month = mar,
          eid = {arXiv:2603.16647},
        pages = {arXiv:2603.16647},
          doi = {10.48550/arXiv.2603.16647},
archivePrefix = {arXiv},
       eprint = {2603.16647},
 primaryClass = {astro-ph.HE},
       adsurl = {https://ui.adsabs.harvard.edu/abs/2026arXiv260316647P}
}

@ARTICLE{2014Caprioli,
       author = {{Caprioli}, D. and {Spitkovsky}, A.},
        title = "{Simulations of Ion Acceleration at Non-relativistic Shocks. I. Acceleration Efficiency}",
      journal = {\apj},
         year = 2014,
        month = mar,
       volume = {783},
       number = {2},
          eid = {91},
        pages = {91},
          doi = {10.1088/0004-637X/783/2/91},
archivePrefix = {arXiv},
       eprint = {1310.2943},
 primaryClass = {astro-ph.HE},
       adsurl = {https://ui.adsabs.harvard.edu/abs/2014ApJ...783...91C}
}

@ARTICLE{1992Cabrit,
       author = {{Cabrit}, Sylvie and {Bertout}, Claude},
        title = "{CO line formation in bipolar flows. III. The energetics of molecular flows and ionized winds.}",
      journal = {\aap},
         year = 1992,
        month = jul,
       volume = {261},
        pages = {274-284},
       adsurl = {https://ui.adsabs.harvard.edu/abs/1992A&A...261..274C}
}

@ARTICLE{2002Beuther,
       author = {{Beuther}, H. and {Schilke}, P. and {Sridharan}, T.~K. and {Menten}, K.~M. and {Walmsley}, C.~M. and {Wyrowski}, F.},
        title = "{Massive molecular outflows}",
      journal = {\aap},
         year = 2002,
        month = mar,
       volume = {383},
        pages = {892-904},
          doi = {10.1051/0004-6361:20011808},
archivePrefix = {arXiv},
       eprint = {astro-ph/0110372},
 primaryClass = {astro-ph},
       adsurl = {https://ui.adsabs.harvard.edu/abs/2002A&A...383..892B}
}

@INCOLLECTION{2007Cabrit,
       author = {{Cabrit}, Sylvie},
        title = "{Jets from Young Stars: The Need for MHD Collimation and Acceleration Processes}",
    booktitle = {Lecture Notes in Physics, Berlin Springer Verlag},
         year = 2007,
       editor = {{Ferreira}, Jonathan and {Dougados}, Catherine and {Whelan}, Emma},
       volume = {723},
        pages = {21},
       adsurl = {https://ui.adsabs.harvard.edu/abs/2007LNP...723...21C}
}

@ARTICLE{1996Bontemps,
       author = {{Bontemps}, S. and {Andre}, P. and {Terebey}, S. and {Cabrit}, S.},
        title = "{Evolution of outflow activity around low-mass embedded young stellar objects}",
      journal = {\aap},
         year = 1996,
        month = jul,
       volume = {311},
        pages = {858-872},
       adsurl = {https://ui.adsabs.harvard.edu/abs/1996A&A...311..858B}
}

@ARTICLE{2014Dunham,
       author = {{Dunham}, Michael M. and {Arce}, H{\'e}ctor G. and {Mardones}, Diego and {Lee}, Jeong-Eun and {Matthews}, Brenda C. and {Stutz}, Amelia M. and {Williams}, Jonathan P.},
        title = "{Molecular Outflows Driven by Low-mass Protostars. I. Correcting for Underestimates When Measuring Outflow Masses and Dynamical Properties}",
      journal = {\apj},
         year = 2014,
        month = mar,
       volume = {783},
       number = {1},
          eid = {29},
        pages = {29},
          doi = {10.1088/0004-637X/783/1/29},
archivePrefix = {arXiv},
       eprint = {1401.2391},
 primaryClass = {astro-ph.GA},
       adsurl = {https://ui.adsabs.harvard.edu/abs/2014ApJ...783...29D}
}

\end{document}